\documentclass[
preprint,
 amsmath,amssymb,
 aps,
]{revtex4-2}

\usepackage{graphicx}
\usepackage{dcolumn}
\usepackage{bm}
\usepackage{mathtools}
\usepackage{amsmath}
\usepackage{ragged2e}
\usepackage{longtable}
\usepackage[usenames,dvipsnames]{color}
\usepackage{float}
\usepackage{siunitx}
\usepackage{accents}
\newcommand{\cbar}[1]{\mkern2mu\overline{\mkern-2mu#1\mkern-1mu}\mkern1mu}

\begin{document}


\title{Tunable Octdong and Spindle-Torus Fermi Surfaces in Kramers Nodal Line Metals}

\author{Gabriele Domaine\textsuperscript{1}}


\author{Moritz M. Hirschmann\textsuperscript{2,3}}

\author{Kirill Parshukov\textsuperscript{2}}

\author{Mihir Date\textsuperscript{1,4}}
\author{Matthew D. Watson\textsuperscript{4}}

\author{Sydney K. Y. Dufresne\textsuperscript{1}}
\author{Shigemi Terakawa\textsuperscript{1,5,6}}
\author{Marcin Rosmus\textsuperscript{7}}
\author{Natalia Olszowska\textsuperscript{7}}


\author{Stuart S. P. Parkin\textsuperscript{1}}

\author{Andreas P. Schnyder\textsuperscript{2}}

\author{Niels B. M. Schröter\textsuperscript{1}}
\email{niels.schroeter@mpi-halle.mpg.de}
\affiliation{\textsuperscript{1}Max Planck  Institut f\"ur  Mikrostrukturphysik, Weinberg 2, 06120 Halle, Germany}

\affiliation{\textsuperscript{2}Max-Planck-Institut für Festkörperforschung, Heisenbergstrasse 1, D-70569 Stuttgart, Germany}

\affiliation{\textsuperscript{3}RIKEN Center for Emergent Matter Science, Wako, Saitama 351-0198, Japan} 

\affiliation{\textsuperscript{4}Diamond Light Source Ltd, Harwell Science and Innovation Campus, Didcot, OX11 0DE, United Kingdom}

\affiliation{\textsuperscript{5}Department of Applied Physics, Graduate School of Engineering, Osaka University, Osaka
565-0871, Japan}

\affiliation{\textsuperscript{6}Center for Future Innovation, Graduate School of Engineering, Osaka University, Osaka
565-0871, Japan}

\affiliation{\textsuperscript{7}SOLARIS National Synchrotron Radiation Centre, Jagiellonian University, Krakow, Poland}

\date{\today}

\begin{abstract}
It has recently been proposed that all achiral non-centrosymmetric crystals host so-called Kramers nodal lines, which are doubly degenerate band crossings connecting time-reversal invariant momenta in the Brillouin zone that arise due to spin-orbit coupling. 
When Kramers nodal lines intersect the Fermi level, they form exotic three-dimensional Fermi surfaces such as Octdong configurations, where all electrons at the Fermi level are two-dimensional massless Dirac fermions. These Fermi surfaces are predicted realize an enhanced form of graphene-like physics in a bulk material, including quantized optical conductivity and giant light-induced anomalous Hall effects. However, until now, no Kramers nodal line metal with such unconventional Fermi surfaces has been experimentally observed. 
Here, we extend the search for Kramers nodal line metals beyond the previously considered case in which the Fermi surfaces enclose a single time-reversal invariant momentum. Using angle-resolved photoelectron spectroscopy measurements and ab-initio calculations, we present evidence that the 3R polytypes of TaS$_2$ and NbS$_2$ are Kramers nodal line metals with open Octdong and Spindle-torus Fermi surfaces, respectively. We show that by reducing the band filling, a transition between these two configurations can be observed. Moreover, our data suggests a naturally occurring size quantization effect of inclusions of 3R-TaS$_2$ in commercially available 2H-TaS$_2$ crystals, which could enable the observation of quantized optical conductivity. Finally, since the open Fermi-surfaces encircle two time-reversal invariant momenta each, we predict a phase transition from a Kramers nodal line metal to a conventional metal by strain or uniaxial pressure. Our work establishes the 3R phase of metallic transition metal dichalcogenides as tunable materials platform to explore new phenomena expected from exotic Fermi surfaces in Kramers nodal line metals.

\end{abstract}

\maketitle


\section{\label{sec:level1}Introduction\protect}
The discovery of graphene revolutionized condensed matter physics by unveiling exotic phenomena such as quantized optical conductivity~\cite{nair2008fine} and light-induced anomalous Hall effects~\cite{mciver2020light}, which arise from the unique two-dimensional Dirac fermions on its Fermi surface. A key limitation of graphene, however, is the presence of only two Dirac fermions at the same energy, which significantly restricts the versatility and magnitude of the aforementioned phenomena. A major challenge in this field has thus been to extend these unique properties in materials with multiple Dirac fermions. Although the discovery of three-dimensional topological semimetals exhibiting multiple Weyl, Dirac or multi-fold fermions represented a significant milestone in that direction~\cite{RevModPhys.90.015001}, these materials typically do not exhibit graphene-like physics, such as quantized optical conductivity~\cite{LinearMartinez2019} due to the three-dimensional nature of their relativistic fermions. Additionally, they pose additional challenges for practical applications since the Dirac and Weyl cones can be located far from the Fermi level.

Recently, it has been predicted that all non-centrosymmetric achiral materials host so-called Kramers nodal lines (KNLs), which are doubly degenerate band crossings connecting time-reversal invariant
momenta (TRIMs) in the Brillouin zone and arise due to the combination of time-reversal symmetry and achiral little group symmetries \cite{Hirschmann_2021,Xie_2021}. When Kramers nodal lines are crossing the Fermi-level they form a Kramers nodal line metal (KNLM), characterized by exotic Octdong (figure-eight) and Spindle-torus Fermi-surfaces, where all electrons are described by a family of two-dimensional Dirac or Rashba Hamiltonians, respectively. Crucially, KNLM arise due to the presence of SOC and have been predicted to lead to fascinating new phenomena ~\cite{Xie_2021}  (Fig. 1a): When the materials are confined, size quantization can lead to a quantized optical conductivity with multiple quantization steps, arising from the presence of multiple Dirac cones at different energies, in contrast to the single quantized value in graphene. Furthermore, one of the Dirac cones is pinned at the Fermi level, meaning that the onset frequency for the quantization is guaranteed to be zero. This is unlike the case of generic Dirac and Weyl points in other materials, and hence lead to a finite onset frequency. Moreover, under light illumination, giant anomalous Hall effects are expected to occur due to the large number of two-dimensional massless Dirac fermions forming the Fermi-surface. Finally, since KNLs are transformed into Kramers-Weyl fermions upon breaking of the mirror or roto-inversion symmetries, KNLM are the parent state for structurally chiral Kramers-Weyl semimetals, which have so far remained elusive in experiments. As a result, KNLM could realize strain induced spin-hedgehogs ~\cite{Krieger_2024}, orbital angular momentum monopoles ~\cite{Yen_2024}, unconventional spin-orbit torques that could be exploited for novel spintronic devices ~\cite{he2021kramers}, and a quantized circular photogalvanic effect without multiband corrections ~\cite{de_Juan_2017}. However, due to the lack of material candidates, these theoretical predictions have so far remained untested experimentally.

It has previously been shown that if the Fermi surface (FS) encloses only one of the TRIMs connected by the KNL, the material is guaranteed to form a KNLM and the touching FSs can be ascribed to either an octdong or spindle-torus ~\cite{Xie_2021}. In the former (latter), two FS pockets that enclose different TRIMS (the same TRIMs) are touching and all electrons on the FS can be described by a family of Dirac (Rashba) Hamiltonians parametrized by the momentum along the nodal line.

Although several materials have been suggested to host KNLs away from the Fermi level, such as SmAlSi~\cite{Zhang_2023} or transition metal ruthenium silicides ~\cite{Shang_2022}, to the best of our knowledge, a KNLM where the KNLs are crossing the Fermi-level has not been conclusively identified in experiments yet. Whilst there have been investigations of KNLMs in the charge density wave state of RTe$_3$ ~\cite{sarkar2023charge,sarkar2024kramersnodallinecharge}, the crossing and splitting of the nodal line at the Fermi level is not resolved in the experiments. Moreover, although there have been predictions for the realization of a spindle-torus FS is real materials, a candidate material hosting an exotic octdong FS that could give rise to quantized optical conductivity has not yet been identified theoretically. In this work, by combining density functional theory (DFT) and angle-resolved photoelectron spectroscopy (ARPES), we find the non-centrosymmetric rhombohedral phases (3R) of  metallic \textit{M}S$_2$ (\textit{M}=Nb, Ta) as candidates for highly tunable KNLMs, realizing both octdong and spindle-torus FSs.

\begin{figure}[H]
    \centering
    \includegraphics[width = 1\textwidth]{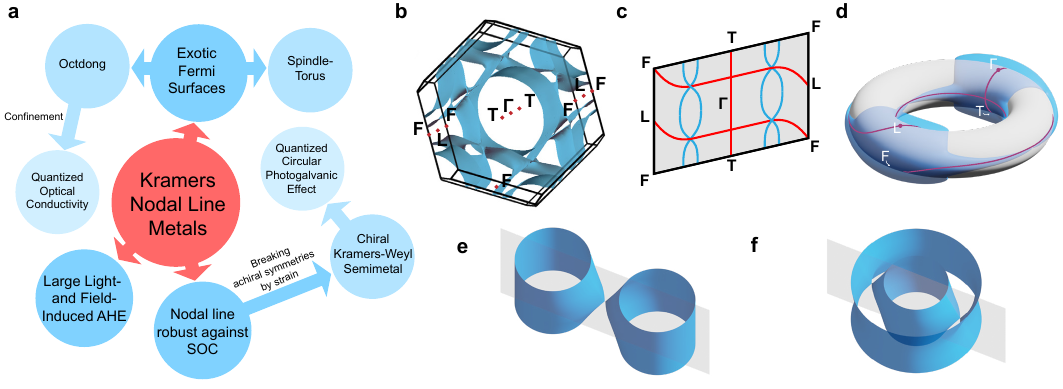}
   \caption{\textbf{Introducing KNLMs.} a) New phenomena arising from KNLMs. b) Fermi-surface computed by DFT of the 3R phase of TaS\textsubscript{2} with  TRIMs indicated by red dots. c) One of the three MIMPs delimited by the high-symmetry point $F$, where an example of the KNL connectivity is shown in red, together with the points of the FS intersecting the MIMP of 3R-TaS$_2$ shown in light blue. Notice that one KNL is pinned to the high symmetry direction $\mathit{\Gamma}-T$, while the almost-movable KNL is connecting $L$ to $F$ along an arbitrary path. d) The quotient space of the MIMP, where the opposite sides of the Brillouin zone are identified, is a torus (shown in grey), while the individual FS pockets that intersect the MIMP (shown in blue) are 2-Tori $T^2$. If the FSs are pierced by a KNL, a touching point is enforced between the tori. Notice that, for simplicity, only the pockets intersecting the MIMP are shown. e) Open octdong FS, formed by an enforced touching between two pockets enclosing different TRIMs, and f) open spindle-torus FS, with an enforced touching of two pockets enclosing the same TRIMs. }
    \label{figure_1}
\end{figure}

For the non-centrosymmetric space group $R3m$ (no.\ 160) of the 3R transition metal dichalcogenides (TMDCs), each mirror-reflection invariant momentum plane (MIMP) supports two KNLs, protected by both time-reversal symmetry and the mirror symmetry, connecting $\mathit{\Gamma}$ to $T$ as well as $L$ to $F$ \cite{Xie_2021} (see also Table~\ref{tab:AMNL} in the Supplementary material). While the former is pinned along the high symmetry path connecting the two TRIMs, the latter is only pinned at $L$ and $F$ 
but is otherwise free to disperse on the MIMP \cite{Hirschmann_2021}, hence the name almost movable KNL (AMKNL). The typical FS of a metallic 3R-TMDC, exemplified by 3R-TaS$_2$, together with an illustration of the nodal lines are shown in Fig.~\ref{figure_1}b and Fig.~\ref{figure_1}c, while the representation of the MIMP torus, obtained by identifying the points $\mathbf{k} \sim \mathbf{k} + \mathbf{G}$, is shown in Fig.~\ref{figure_1}d.
Although in TMDCs the FSs typically enclose more than one single TRIM, depending on the dispersion of the AMKNL, the realization of a KNLM is still possible. The touching FSs are then realizing an open version of the octdong (Fig.~\ref{figure_1}e) or open spindle-torous (Fig.~\ref{figure_1}f) that are described by the same type of Hamiltonians as their closed counterparts, and therefore lead to equivalent physical responses.

Interestingly, as we predict theoretically and demonstrate experimentally in this work, the FSs of TMDCs can undergo a Lifshitz transition from octdong and spindle-torus induced by a change of the band filling, which can be also be achieved by doping or gating. In particular, our DFT calculations predict that due to the lower band filling, 3R-TaS$_2$ is hosting open octdong Fermi-surfaces. On the other hand, due to a higher band filling, NbS$_2$ realizes open spindle torus FSs where the two touching Fermi-surfaces enclose the same TRIMs 
(Fig.~\ref{figure_1}c), $\mathit{\Gamma}$ and $T$. We confirm these predictions experimentally by directly visualizing these FSs by ARPES.

While the 2H-phase of \textit{M}S$_2$ can be synthesized in stoichiometric single crystals, the 3R-phase has been found to require doping and intercalation to be stable at ambient conditions \cite{Gotoh_1998,El_Youbi_2021}. 
This presents a formidable challenge for experimental confirmation of KNLs, because doping and intercalation can lead to broadening of experimental ARPES spectra. However, it is well-known that \textit{as grown} single crystals of hexagonal transition metal dichalcogenides, in particular transition metal disulfides, are susceptible to stacking faults~\cite{Meetsma_1989}, 
which can lead to locally distinct stacking configurations and the realization  of a 3R-stacking configuration without the need for additional doping. Some of the present authors have recently used microfocused ARPES to obtain relatively sharp spectra of few layer 3R-stacking configurations that were embedded within a single crystal of the 2H-polytype~\cite{El_Youbi_2021}. 
Here,  we make use of these new capabilities to access small domains of 3R-TaS$_2$ with very sharp bands suggesting a low intercalation level for ARPES measurements, which enables us to confirm the presence of octdong Fermi-surfaces in this compound. We furthermore present ARPES measurements of Nb self-intercalated 3R-NbS$_2$, which shows that the measured Fermi-surface is consistent with a spindle-torus FS. Our DFT calculations show that one can tune between those two Fermi-surface types by changing the band filling, which can in principle be achieved by chemical doping or gating. Based on a tight-binding model, 
we finally demonstrate that due to the fact that the FSs enclose multiple TRIMs, applying pressure or strain can transform these materials from Kramers-nodal line metals into ordinary metals, where the KNL does not pierce the FS, by changing the topology of the AMKNL.

\newpage

\section{\label{sec:level2}Results\protect}

\subsection{\label{sec:sublevel2}KNLM with octdong FS in 3R-TaS$_2$}

\begin{figure}[H]
    \centering
    \includegraphics[width = 1\textwidth]{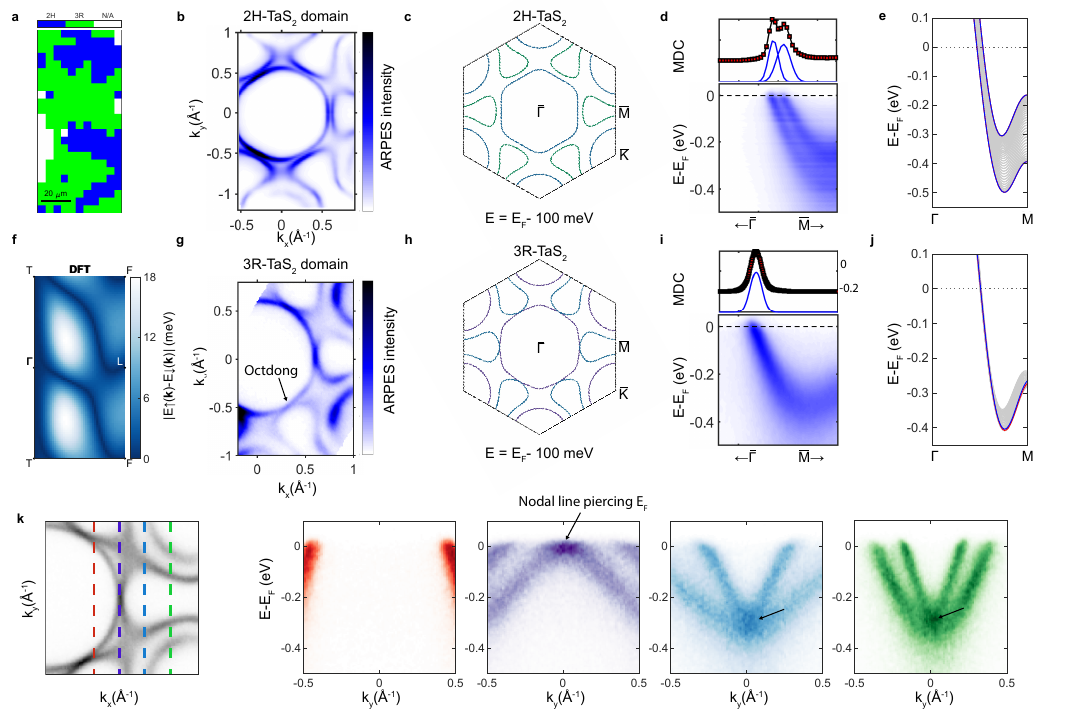}
    \caption{\textbf{Octdong FS formed by nodal line piercing E$_F$ in 3R-TaS$_2$.} \textbf{a}) Spatial map of a cleaved sample showing the presence of different domains identified by the number of bands crossing the Fermi level along the MIMP ($\cbar{\mathit{\Gamma}}-\cbar{M}$ direction).  \textbf{b}) FS of 2H-TaS$_2$ measured with h$\nu =$ 50 eV and LH (p) polarized light and \textbf{c}) the corresponding FS at the $\mathit{\Gamma}$ plane computed by DFT. 
    \textbf{d}) ARPES data of the 2H phase along the MIMP and \textbf{e}) corresponding  DFT calculation for $k_z = 0$ ($k_z \neq 0$) in blue (gray). The upper panel of \textbf{d}) is the corresponding momentum distribution curve taken at the Fermi level. \textbf{f}) DFT calculation of the band splitting between the two bands crossing the Fermi level on the MIMP in 3R-TaS$_2$. Both the KNL pinned along $\mathit{\Gamma}-T$ as well as the AMKNL are visible. g) FS of 3R-TaS$_2$ measured with h$\nu =$ 55 eV and and LH (p) polarized light and \textbf{h}) the corresponding FS at the $\mathit{\Gamma}$ plane computed by DFT. \textbf{i}) ARPES band structure of the 3R phase along the MIMP and \textbf{j}) corresponding DFT calculations for $k_z = 0$ ($k_z \neq 0$) showing the split bands in blue and red (gray). The upper left panel of i) is the corresponding momentum distribution curve taken at the Fermi level. 
    Panels \textbf{k}) show the dispersion along three different directions perpendicular to the MIMP on both sides of the octdong touching point. While going from the central hole pocket to the external electron pocket the Dirac crossing transitions from above (red) to below (blue and green) the Fermi level, as predicted for the octdong FS in Ref.~\cite{Xie_2021}.}
    \label{figure_2}
\end{figure}

Because both SOC and non-centrosymmetric space groups are required for the realization of a KNLM, TMDCs involving transition metals, like Nb or Ta, are ideal candidates. During our micro-ARPES measurements of  commercially available cleaved TaS$_2$ single crystals nominally in the 2H phase, besides the typical FS of the 2H phase, we often observed variably sized domains of another phase exhibiting one single band crossing (within experimental resolution) at the Fermi level along the $\mathit{\Gamma}-M$ direction, which we identify as the 3R phase. The characteristic sizes of their respective domains are shown in Fig. \ref{figure_2}a, where the distinction between 3R, 2H, and any other possible stacking can be made conclusively by counting the number of bands at each spot as explained in the supplementary material section \ref{supplementary_Spatial_map}. The main differences between the band structures of the two phases are highlighted in Fig. \ref{figure_2}. 

The measured FS of a 2H-TaS$_2$ domain is shown in Fig.~\ref{figure_2}a, together with the corresponding density functional theory (DFT) calculation in Fig.~\ref{figure_2}b. The experimental FS shows a central hole pocket surrounded by three 'dog-bone' electron pockets along the $\cbar{\mathit{\Gamma}}-\cbar{M}$ direction (along the MIMP). Importantly, the two pockets are not connected, as can also be seen from the experimental band dispersion along $\cbar{\mathit{\Gamma}}-\cbar{M}$ displayed in Fig.~\ref{figure_2}d, which is clearly showing two bands crossing the Fermi level. These two bands are spin-degenerate because of the presence of time-reversal- and inversion-symmetry and are mostly split by interlayer coupling, which is also confirmed by our DFT calculations shown in Fig.~\ref{figure_2}e. 

In contrast, the experimental Fermi-surface of the 3R phase (Fig.~\ref{figure_2}g) shows that the two pockets are connected into a degenerate point, forming a FS shaped like an hourglass. Only a single band is resolvable in the band dispersion along $\cbar{\mathit{\Gamma}}-\cbar{M}$ in Fig. \ref{figure_2}i, while two bands are still visible from the experimental Fermi-surface away from the MIMP. Our DFT calculations of 3R-TaS$_2$, shown in Fig.~\ref{figure_2}h and Fig.~\ref{figure_2}j, reproduce the same features observed by ARPES, strongly supporting the identification of the 2H and 3R polytypes.

This implies that the hourglass touching point along the MIMP, if pierced by the AMKNL, can be ascribed to the open octdong formed by joining together the 'dog-bone' electron pocket to the central hole pocket. Our DFT calculations (Fig.~\ref{figure_2}f) predict a nodal line that is fully winding around the MIMP torus, implying that the nodal line must pierce the FS. By comparing three band dispersions along a direction perpendicular to the MIMP (Fig.~\ref{figure_2}k-n), the experimental data clearly reveals the Kramers nodal line crossing the Fermi level (Fig.~\ref{figure_2}m). The Dirac-like dispersion that is shifting through the Fermi level when going from touching hole to electron pockets is the predicted hallmark of the octdong FS of KNLMs and has never been observed in experiments (c.f. Fig. 3e in Ref. ~\cite{Xie_2021}). Note that the gapless nodal line only occurs at specific out of plane momenta k$_z$ within the MIMP, which would technically correspond to a specific photon energy in an ARPES experiment. However, due to the limited probing depth of our experiment that results in out-of-plane momentum broadening ~\cite{strocov2003intrinsic}, we observe a gapless nodal line spectrum for all photon energies within our experimental resolution. 

\subsection{\label{sec:sublevel2}Naturally occurring quantum confinement in 3R-TaS$_2$ samples}
\begin{figure}[h]
    \centering
    \includegraphics[width = 1\textwidth]{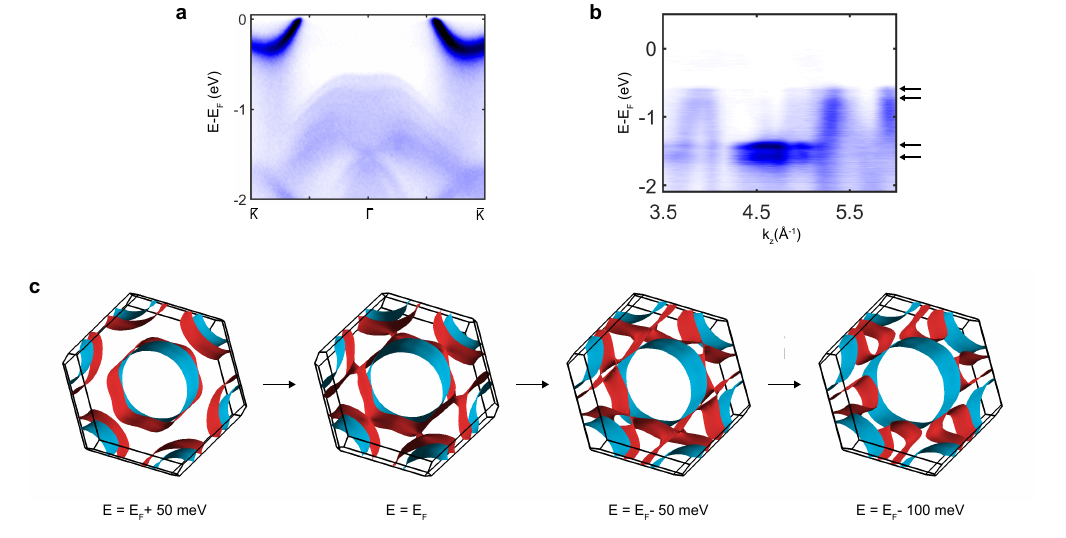}
    \caption{\textbf{Quantum-well states observed in 3R-TaS$_2$ and Fermi-surface transition with band filling. }a) band structure of 3R-TaS$_2$ at 72 eV on the MIMP, showing a quantization of the valence band. b) Quasi-two-dimensional character of some of the bands at $k_x = 0$ as shown by their flat dispersion as a function of the out-of-plane momentum $k_z$. c) Change in the topology of isoenergy surfaces of 3R TaS$_2$ upon change of the binding energy. As the binding energy  is reduced, the system transitions from concentric pockets to pockets surrounding different TRIMs.}
    \label{figure_5}
\end{figure}

Identifying a naturally occuring quantum confinement of the 3R-TaS$_2$ within the 2H-TaS$_2$ sample would be particularly appealing since the octdong FS can be described by a two-dimensional Dirac Hamiltonian, which leads to a quantization of the optical conductivity when the momentum along the nodal line is quantized. This was originally proposed by considering size quantization in the thin film limit \cite{Xie_2021}.  Intriguingly, our ARPES measurements of the 3R-TaS$_2$ domains show signatures of quantum well states in the valence band (Fig.~\ref{figure_5}a and Fig.~\ref{figure_5}b), similar to the 3R-NbS$_2$ and 3R-MoS$_2$ thin film inclusions that were recently reported in commercial single crystals nominally of the 2H-polytype \cite{Watson_2024}. This observation suggests that the 3R phase on the surface of the commercially available 2H-TaS$_2$ could consist of only a few layers, which would also lead to a quantization of the out-of-plane momentum, implying that it could directly exhibit a quantization of the optical conductivity. An alternative explanation of our data could be surface band bending, which has also been invoked to explain quantum well states in the valence band of TaSe$_2$ \cite{Li_2021},
due to a confining surface potential. Such band-bending induced surface confinement would also lead to a quantization of the conduction band and thus quantized optical conductivity similar to the thin film case, which could be investigated in future optical experiments. A third explanation could be that the observed valence band features are actually genuine surface states or surface resonances rather than quantum well states, in which case the conduction band would not have to be quantized.

\subsection{\label{sec:sublevel2}Tuning the octdong into a spindle-torus by increased band filling in 3R-TMDCs}
Another interesting feature of the 3R-TMDCs is that, because of their characteristic band structure along the $\cbar{\mathit{\Gamma}}-\cbar{K}$ path, increasing the filling factor can lead to a Lifshitz transition from the open octdong to the open spindle-torus configuration, as shown by the DFT calculation comparing isoenergy surfaces at different binding energy of 3R-TaS$_2$ in Fig. \ref{figure_5}c. Since only the octdong FS is expected to show quantized optical conductivity, doping or gating these compounds could be used to significantly tune the optoelectronic properties of these materials. To demonstrate this tunability with band filling, we also investigated the commercially available 3R-NbS$_2$ polytype, which is known to have an increased band-filling due to self-doping by Nb interstitials during growth ~\cite{El_Youbi_2021}. This material is predicted by DFT to exhibit two touching concentric FSs centred at $\mathit{\Gamma}$ (Fig.~\ref{figure_3}a), surrounded by six disconnected FS pockets that are split due to spin-orbit coupling. The predicted FS is also confirmed by ARPES, showing a central hole pocket surrounded by six disconnected pockets (Fig.~\ref{figure_3}b). 
Since the spin-orbit coupling in this compound is relatively small and the spectral features are broad, likely due to the presence of interstitial Nb atoms~\cite{El_Youbi_2021}, we were not able to resolve the predicted spin-orbit splitting of the bands in our ARPES measurement. However, our DFT calculations predict an almost movable nodal line that fully winds twice around the MIMP torus, so that it is guaranteed to cross the FS at four points. This would imply that the two hole pockets enclosing $\mathit{\Gamma}$ and $T$ realize an open spindle-torus FS, which is consistent with our ARPES measurements.

\begin{figure}[h]
    \centering
    \includegraphics[width = 1\textwidth]{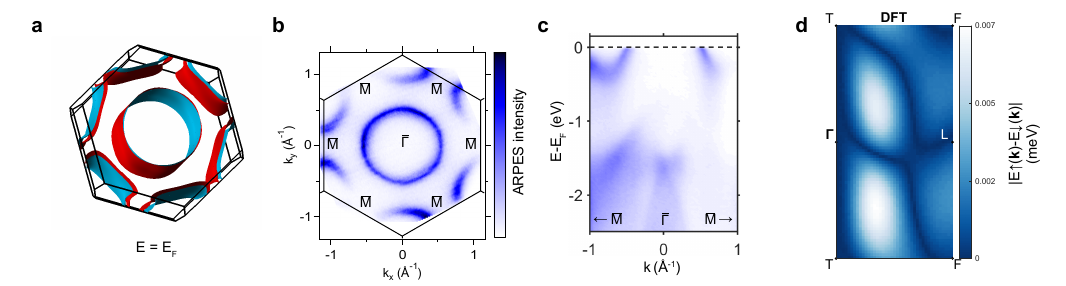}
    \caption{\textbf{Spindle-torus FS in 3R-NbS$_2$. }a) Three-dimensional FS of 3R-NbS$_2$ computed by DFT. b) 70 eV ARPES FS of 3R-NbS$_2$ and c) band dispersion along the high-symmetry path $\cbar{\mathit{\mathit{\Gamma}}}-\cbar{M}$. d) Magnitude of the splitting between the two bands crossing the Fermi level on the MIMP, showing the AMKNL connecting $L$ to $F$ by winding around the MIMP torus.}
    \label{figure_3}
\end{figure}

\subsection{Nodal line topology and its tunability by strain}


As already mentioned in the previous sections, while the existence of the nodal lines is symmetry-enforced, the dispersion of the AMKNL is material-dependent so that one can make a distinction between those that are guaranteed to cross the FS and those that are not.
To do so, one can start by realizing that the quotient space of the MIMP with the prescription $\mathbf{k} \sim \mathbf{k} + \mathbf{G}$, where  $\mathbf{G}$ is a reciprocal lattice vector, is a torus $T^2_{MIMP}$. On the other hand, both the mirror-invariant points of the FS and the AMKNL connecting $L$ and $F$ can be ascribed to one dimensional loops $S^1 \subset T^2_{MIMP}$. The different mappings $f$ of loops onto $T^2$ are elements of the fundamental homotopy group $\pi_1(T^2) = \mathbb{Z} \times \mathbb{Z}$ where each class is identified by a set of two winding numbers, respectively along the meridian and longitude, $f \in [w_m,w_l] \in \pi_1(T^2)$. If we take the high-symmetry path $\mathit{\Gamma}-T$ to lie along the torus' meridian, then we have that the FS loops (FSL) and the pinned KNL connecting $\mathit{\Gamma}$ and $T$ both belong to the classes [$\pm$1,0]. Because for symmorphic mirror symmetries there can only be an odd number of lines originating from each TRIM due to TRS exchanging the mirror symmetry eigenvalues, the AMKNL is restrained to the classes [2$\mathbb{Z} + 1$,2$\mathbb{Z}$]. Then, if $f_{AMKNL} \in [1,w_l \neq 0]$, each FSL must be pierced by the nodal line at least $w_l$ times. On the other hand, if $f_{AMKNL} \in [w_m,0]$ the crossing is not guaranteed and will depend on the relative position of the nodal line and the FS. This means that there can be two types of KNLM, those with $w_l \neq 0$ where the FS crossing is guaranteed, and those with $w_l=0$, which are accidental (Fig. \ref{figure_5}).
Because each $w_l$ identifies topologically distinct nodal lines, a transition from one class to the other must happen through a Lifshitz transition of the AMKNL. 
For both NbS$_2$ and TaS$_2$ the DFT calculations predict $w_l=2$, however we expect it to be possible to enforce a transition to $w_l=0$ by applying strain to the system to affect the out-of-plane dispersion of the nodal line. To provide an intuitive understanding of this, we construct a tight-binding model with a single orbital and a spin-1/2 degree of freedom.
The parameters are the on-site energy $t_0$, all nearest-neighbour terms $t_2$ up to $t_5$, and all second-nearest-neighbour hopping terms, $t_6$ and $t_7$.
For nearest-neighbour hopping, the spin-orbit splitting within the MIMPs, and thus the shape of the nodal lines, is fully determined by the relative sizes of out-of-plane hopping $t_2$ and the in-plane hopping $t_3$. 
When the in-plane hopping dominates (Fig.~\ref{figure_5}a), the spin-orbit gap depends mostly on the in-plane momentum, and thus only closes when traversing the Brillouin zone in the in-plane direction. If instead the out-of-plane hopping is larger (Fig.~\ref{figure_5}b), the gap closes along the vertical direction, hence the nodal line crosses the $\mathit{\Gamma}-T$ path. 
Since the corresponding hopping terms, $t_2$ and $t_3$, are momentum-dependent, compressive strain increasing the out-of-plane hopping, and thus induces a transition of nodal line connectivity within our tight-binding model. 
This model behavior is robust, because the only other nearest-neighbor spin-orbit splitting $t_5$ vanishes on the MIMPs. 
Adding the second-nearest-neighbor hopping introduces the out-of-plane spin-orbit term, $t_7$, which, like $t_2$, favors a nodal line crossing $\mathit{\Gamma}$-T. So the expected effect of compressive strain does not change under the addition of longer-range hopping.  Details of the complete Hamiltonian can be found in the Methods \ref{tight_binding_methods}.


\begin{figure}[H]
    \centering
    \includegraphics[width = 1\textwidth]{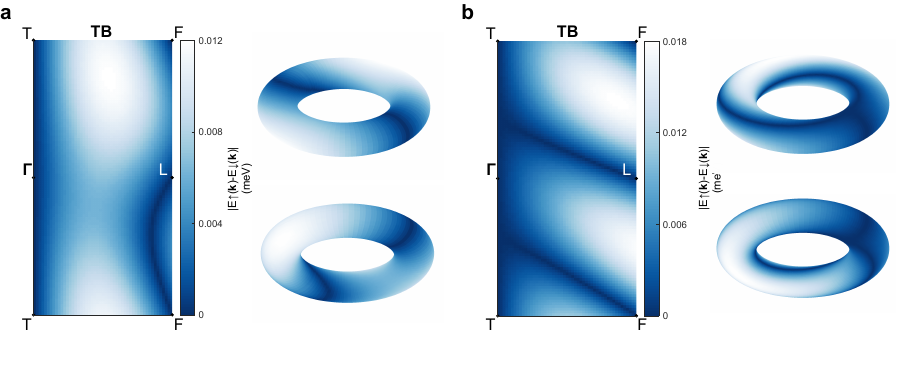}
    \caption{\textbf{Tunability of nodal line connectivity with strain. }a-b) Splitting between the two bands crossing the Fermi level showing two different types of connectivity. In a) the in-plane hopping that is stronger (b) (weaker) than out-of-plane hopping, $t_2 = 0.3$~meV, $t_3 = 1.3$~meV ($t_2 = 1.3$~meV, $t_3 = 0.3$~meV). The figures also show the mapping of the nodal lines onto the MIMP torus (respectively viewed from the top and bottom), where it is apparent that the configuration with the stronger in-plane hopping does not wind around the torus, and as such is not guaranteed to cross the FS. On the other hand, the configuration with a stronger out-of-plane hopping fully winds twice around the torus, and is thus guaranteed to cross the FS. Notice also that the nodal line winding along the meridian on the right-hand side of the torus corresponds to the pinned KNL along $\mathit{\Gamma}-T$.}
    \label{figure_3}
\end{figure}

\section{Discussion and conclusion}

We have demonstrated using ARPES and theoretical calculations that the 3R polytypes of TaS$_2$ and NbS$_2$ are KNLMs, exhibiting, respectively, octdong and spindle torus FSs enforced by a KNL crossing the Fermi level. We have furthermore demonstrated that FSs that enclose multiple TRIMs allow for high tunability between octdong and spindle-torus by band filling, as well as topological KNLM to trivial metal transition by strain. Future experiments could test the quantized optical conductivity of confined samples with octdong FS of 3R-TaS2, which we have demonstrated to naturally occur on the surface of as grown 2H-TaS2 due to stacking faults or surface band bending. 
Moreover, the gate-tunability of these systems could even allow for junctions between octdong and spindle-torus FSs that could be used to investigate the scattering of Dirac- into Rashba-electrons, and vice-versa. 
These findings establish the 3R polytypes of metallic TMDCs as an ideal tunable platform for studying new phenomena predicted for KNLMs.

\section{Methods}

\subsection{ARPES}
Single crystals of 3R-NbS$_2$ and 2H-TaS$_2$ were obtained commercially from HQ Graphene, Groningen.
ARPES measurements of TaS$_2$ were performed at the I05 beamline of Diamond Light Source \cite{Hoesch_2017}, with a Scienta DA-30 analyser measaured at a pressure of approx. 1-2 x 10\textsuperscript{-10} mbar, and using photon energies in the range 30–150 eV, at sample temperatures between 80K and 100K. The combined energy resolution was approximately 30-40 meV with an approximate beam spot diameter (FWHM) of 4 $\mu m$.
On the other hand, the measurements of NbS$_2$ were performed at the URANOS beamline at the National Synchrotron Radiation Centre SOLARIS in Krakow (Poland) using a SCIENTA OMICRON DA30L photoelectron spectrometer, at the sample temperature of 15K. The photon energy used ranged from 20-100 eV.

\subsection{DFT}

Band structures and FSs of (2H, 3R)-TaS$_2$ and 3R-NbS$_2$ were obtained with DFT. We used Perdew–Burke–Ernzerhof functional implemented in the Vienna Ab Initio Simulation Package (VASP)~\cite{PBE,VASP_1,VASP_2,VASP_3}  utilizing projector-augmented wave method~\cite{VASP_PAW,PAW} with kinetic energy cutoff for the plane-wave basis 520 eV. Both 3R phases are described by symmetries of SG 160 (R3m) with atoms located at 3a Wyckoff position. Optimized positions of atoms for 3R-NbS$_2$ were taken from Materials Project~\cite{Materials_Project} with lattice parameters $a=b=c=7.087 \ \si{\angstrom}, \ \alpha=\beta=\gamma=27.403^{\circ}$, the compound has the ID mp-966 (ICSD 42099~\cite{ICSD_database}). For the calculation of 2H-TaS$_2$ (SG 194) electronic structure, we relaxed a hexagonal cell with lattice parameters $a=b=3.342\ \si{\angstrom},\ c=13.760\ \si{\angstrom},\ \alpha=\beta=90^{\circ},\mathit{\Gamma}=120^{\circ}$ until the forces on all atoms were less than $1e^{-6}$. The compound has the IDs mp-1984 and ICSD 68488. Relaxed positions of atoms for 3R-TaS$_2$ were taken for the compound with the code mp-10014 (ICSD 43410), the lattice parameters were  $a=b=c=7.089 \ \si{\angstrom}, \ \alpha=\beta=\gamma=27.281^{\circ}$. To compute the FS we performed the calculations on a k-grid $41 \times 41 \times  41$. 

\subsection{Tight binding model}\label{tight_binding_methods}

The complete Hamiltonian is expressed as 
\begin{align}
    H(\bm{k})
    &= 
    \sum_{n= 1}^7 H_n(\bm{k}),
\end{align}
where, using the primitive lattice vectors $\bm{a}_1,\bm{a}_2,\bm{a}_3$ and a spin-1/2 basis, the matrices $H_n(\bm{k})$ are given as
\begin{align}
    H_1(\bm{k}) &= (t_1 + t_1^*) 1_{2\times2} ,
    \\
    (H_2(\bm{k}))_{12} &= (H_2(\bm{k}))^*_{21} 
    \nonumber\\
    &=
    2i (t_2 + t_2^* \mathrm{e}^{i \frac{\pi}{3} } ) \left( 
    \mathrm{e}^{ i \frac{2\pi}{3} } \sin( \bm{k} \cdot \bm{a}_1)
    +\sin( \bm{k} \cdot \bm{a}_2)
    +\mathrm{e}^{- i \frac{2\pi}{3} }  \sin( \bm{k} \cdot \bm{a}_3)
    \right) ,
    \\
    (H_3(\bm{k}))_{12} &= (H_3(\bm{k}))^*_{21} 
    \nonumber\\
    &=
    2i (t_3 + t_3^* \mathrm{e}^{ i \frac{2 \pi}{3} })
    \big( 
    \mathrm{e}^{ i \frac{2\pi}{3} } \sin( \bm{k} \cdot (\bm{a}_1 - \bm{a}_2))
    + \sin( \bm{k} \cdot (\bm{a}_2 - \bm{a}_3))
    \nonumber\\
     &\qquad\qquad\qquad+ \mathrm{e}^{- i \frac{2\pi}{3} } \sin( \bm{k} \cdot (\bm{a}_3 - \bm{a}_1)) 
    \big) ,
    \\
    (H_4(\bm{k}))_{11} &= (H_4)_{22} 
    \nonumber\\
    &= 2 (t_4 + t_4^*) \left( \cos(\bm{k} \cdot \bm{a}_1 ) + \cos(\bm{k} \cdot \bm{a}_2 ) + \cos(\bm{k} \cdot \bm{a}_3 ) \right) ,
    \\
    (H_5(\bm{k}))_{11} &= (H_5(-\bm{k}))_{22} 
    \nonumber\\
    &=
    2 \Re(t_5) \left( 
    \cos( \bm{k} \cdot (\bm{a}_1 - \bm{a}_2))
    +\cos( \bm{k} \cdot (\bm{a}_2 -\bm{a}_3))
    +\cos( \bm{k} \cdot (\bm{a}_3 - \bm{a}_1)) 
    \right)
    \nonumber\\
    &+2 \Im(t_5)
    \left( 
    \sin( \bm{k} \cdot (\bm{a}_1 - \bm{a}_2)) 
    +\sin( \bm{k} \cdot (\bm{a}_2 - \bm{a}_3))
    +\sin( \bm{k} \cdot (\bm{a}_3 - \bm{a}_1))
    \right) ,
    \\
    (H_6(\bm{k}))_{12} &= (H_6(\bm{k}))^*_{21} 
    \nonumber\\
    &=
    2i (t_6 - t_6^*)
    \big( 
    \mathrm{e}^{+ i \frac{2\pi}{3} } \sin( \bm{k} \cdot (\bm{a}_1  - \bm{a}_2 + \bm{a}_3)) 
    - \sin( \bm{k} \cdot (\bm{a}_1  + \bm{a}_2 - \bm{a}_3))
    \nonumber\\
    &\qquad\qquad\qquad+ \mathrm{e}^{- i \frac{2\pi}{3} } \sin( \bm{k} \cdot (-\bm{a}_1  + \bm{a}_2 + \bm{a}_3))
    \big) ,
    \\
    (H_7(\bm{k}))_{11} &= (H_7(\bm{k}))_{22} 
    \nonumber\\
    &=
    2 (t_7 + t_7^*) \big( 
    \cos(\bm{k} \cdot (-\bm{a}_1 + \bm{a}_2 + \bm{a}_3) )
    + \cos(\bm{k} \cdot (\bm{a}_1 - \bm{a}_2 + \bm{a}_3) )
    \nonumber\\
    &\qquad\qquad\qquad+ \cos(\bm{k} \cdot (\bm{a}_1 + \bm{a}_2 - \bm{a}_3) )
    \big) .
\end{align}
All not further specified terms in the matrices $H_n(\bm{k})$ are zero. 

\section{Acknowledgments}

G.D. and S.K.Y.D acknowledge proposal SI39232 at the i05-1 Endstation at the Diamond Light Source, UK. G.D. and S. T. acknowledge proposal 20227155 at the URANOS beamline at the National Synchrotron Radiation Centre “SOLARIS”, Poland. This publication was partially developed under the provision of the Polish Ministry of Science and Higher Education project 'Support for research and development with the use of research infrastructure of the National Synchrotron Radiation Centre SOLARIS' under contract no 1/SOL/2021/2. N.B.M.S. acknowledges funding by the European Union (ERC Starting Grant ChiralTopMat, Project No. 101117424). M.M.H. is funded by the Deutsche Forschungsgemeinschaft (DFG, German Research Foundation) - project number 518238332. K.P. and A.P.S. are funded by the Deutsche Forschungsgemeinschaft (DFG, German Research Foundation) – TRR 360 – 492547816. G.D. acknowledges support by the Max Planck Graduate Center for Quantum Materials (MPGC-QM). G.D. and M.D. acknowledge Jenny Davern for her assistance in conceptualizing Fig. 1a. 

\bibliographystyle{apsrev4-1}
\bibliography{bibliography-28-08-2024-11}

\section{Supplementary}

\subsection{Almost movable nodal lines}

Almost movable nodal lines (AMNLs) are enforced by crystal symmetries \cite{Hirschmann_2021}, e.g., by those of the 3R phase of TaS$_2$. AMNLs are KNLs~\cite{he2021kramers} that are only pinned to TRIMs, but not to high-symmetry paths. Inside the mirror plane, such a line degeneracy connecting TRIMs can have an arbitrary shape as long as the global connectivity is respected~[see Fig.~\ref{figure_5}(a,b)].

We give the comprehensive classification of space groups enforcing AMNLs in Table~\ref{tab:AMNL}~\cite{masters_thesis_Parshukov}. 
For the AMNL space groups listed in Table~\ref{tab:AMNL}, we also note all other symmetry-enforced band crossings in the presence of SOC, including twofold (Weyl) and fourfold points, pinned nodal lines, and movable hour-glass nodal lines which are denoted following the notation of Ref.~\cite{Hirschmann_2021,Leonhardt_2021}.
AMNL and movable hour-glass nodal lines are denoted by a bracket listing TRIMs and/or degenerate lines on the mirror plane separated by a semicolon. 
At the TRIMs before (after) the semicolon different (identical) mirror eigenvalues are paired by the space group symmetries, e.g., time-reversal symmetry for the AMNL. 
By the argument provided for the existence of AMNL \cite{Hirschmann_2021}, each TRIM on the left side of the semicolon in this notation is part of a nodal line, e.g., for space group (SG) 6 $Pm$ the bracket ($\Gamma$,B,Y,A;--) all TRIMs within the mirror plane are part of AMNLs. 
This is generally true, but for not listed SGs the nodal line passing through such TRIMs is pinned instead of AMNL. 

On any path within the Brillouin zone connecting points with the same mirror eigenvalue pairing, the number of nodal lines must be even, so in the simplest case zero.
Between points with different pairing, there have to be an odd number of nodal lines, to provide the necessary exchange of mirror eigenvalues. 
The latter implies the presence of hour-glass nodal lines. 
Notably, AMNL and hour-glass nodal lines reside within different (direct) band gaps, AMNL occur within all odd numbered gaps, $2n+1, \,\, n \in \mathbb{N}_0$ sorted in energy, whereas hour-glass nodal lines occur within every second even numbered gap, $2+4n$. 
Notably, some TRIMS exhibit AMNL depending on the orbital character of the bands, in other words, the realized symmetry representation. 
In these cases, a pinned nodal line can but does not always exist that crosses a TRIM and replaced the AMNL, we have marked the TRIMs and pinned nodal lines with a star in Table~\ref{tab:AMNL}.

\begin{longtable}{||p{2cm} | p{2.7cm}|p{1.7cm}| p{2.5cm}|  p{6.7cm}||} 
\hline
 SG & AMNL position &  points & other lines &  comment \\
\hline
6 Pm& ($\Gamma$,B,Y,A;--) (Z,C,D,E;--) &  &   &  \\
\hline
7 Pc& ($\Gamma$,Y;--) (Z,C;--) & &  ($\Gamma$,Y;B,A), (Z,C;D,E), BD, AE&  \\
\hline
8 Cm& ($\Gamma$,Y,A,M;--)  & L(2),V(2) &  & \\
\hline
9 Cc& ($\Gamma$,Y;--) &L(2),V(2) & ($\Gamma$,Y;A,M), AM &  \\
\hline
38 Amm2& (S;--) (R;--) &&$\Gamma$Y, ZA&\\
\hline
40 Ama2 & (S;--) (R;--) &&($\Gamma$Y;Z,T), \ ZT&\\
\hline
44 Imm2& (S;--) (R;--)&T(2)&$\Gamma$X&\\
\hline
46 Ima2& (S;--)&T(2)&RW, ($\Gamma$X;R)&\\
\hline
107 I4mm& (N;--)&&$\Gamma$M, PX&\\
\hline
109 I4$_1$md& (N;--) (P$^*$;--)&M(4)&($\Gamma$M;X), XM&AMNLs at P appear for a particular 2-fold rotation eigenvalue\\
\hline
110 I4$_1$cd &(P;--)&M(4)&($\Gamma$M;N)(4), (XM;P)(4)&\\
\hline
119 I-4m2& (N;--)&&$\Gamma$M&\\
\hline
156 P3m1& (M,L,$\Gamma^*$,A$^*$;--)&&$\Gamma$A$^*$&Depending on $\Gamma$A rep, pinned NL or 3 AMNLs in 3 mirror planes at $\Gamma$, A\\
\hline
157 P31m& (M,L,$\Gamma^*$,A$^*$;--)&&$\Gamma$A$^*$, KH$^*$&Depending on $\Gamma$A rep, pinned NL or 3 AMNLs in 3 mirror planes at $\Gamma$, A; NL at KH appears for a particular rep\\
\hline
158 P3c1& (M,$\Gamma^*$;--)&A(4)$^*$&($\Gamma$,M;A,L), AH, LH, $\Gamma$A$^*$&Depending on $\Gamma$A rep, pinned NL or 3 AMNLs in 3 mirror planes at $\Gamma$\\
\hline
159 P31c& (M,$\Gamma^*$;--)&A(4)$^*$&($\Gamma$,M;A,L), AL, KH$^*$, $\Gamma$A$^*$&Depending on $\Gamma$A rep, pinned NL or 3 AMNLs in 3 mirror planes at $\Gamma$\\
\hline
160 R3m& (L,F,$\Gamma^*$,T$^*$;--)&&$\Gamma$T$^*$, P$^*$&Depending on $\Gamma$T rep, pinned NL or 3 AMNLs in 3 mirror planes at $\Gamma$, T; P is symmetric under mirror and 3-fold rotational symmetries, the NL appears for a particular rep\\
\hline
161 R3c& (F,$\Gamma^*$;--)&T(4)$^*$&($\Gamma$,F;T,L), $\Gamma$T$^*$, P$^*$, Y, B&Depending on $\Gamma$T rep, pinned NL or 3 AMNLs in 3 mirror planes at $\Gamma$; P is symmetric under mirror and 3-fold rotational symmetries, the NL appears for a particular rep; Y, B are the $\mathcal{TM}$-symmetric lines\\
\hline
174 P-6& ($\Gamma$,M;--) (A,L;--)&&$\Gamma$A$^*$&At $\Gamma$, A 3 AMNL intersect\\
\hline
216 F-43m& (L;--)$^*$&$\Gamma(4)^*$&$\Gamma$X, $\Gamma$L$^*$&Depending on $\Gamma$L rep, pinned NL or 3 AMNLs in 3 mirror planes at L\\
[1ex] 
\hline
  \caption{\justifying Almost movable nodal lines. All points are given in the Bilbao Crystallographic Server (BCS)~\cite{Aroyo:xo5018} notation. Other lines are taken from Refs.~\cite{Hirschmann_2021,Leonhardt_2021} for tetragonal and orthorhombic SGs, otherwise from the BCS. Mark $^*$ indicates that the existence of the topological crossing depends on the representation. Notation $(\Gamma_i;-)$ represents the fact that the AMNL passes $\Gamma_i$ TRIM. Numbers in brackets show the degeneracy at high-symmetry points and lines. $(A;B)$ represents the hourglass nodal line between degenerate points (or lines)~\cite{Hirschmann_2021,Leonhardt_2021}.}
  \label{tab:AMNL}
\end{longtable}

\subsection{Identification of TaS$_2$ polytypes with ARPES}
TMDCs can crystallize into various polytypic structures~\cite{Katzke_2004}. In the well studied 2H-TaS$_2$ phase, two doubly degenerate bands are crossing the Fermi level along the $\Gamma$-$M$ direction, which are split due to interlayer interaction, and exist due to two metal atoms per unit cell (Fig. \ref{figure_2}b-e). In contrast, the 3R-TaS$_2$ phase, there is only a single band along the $\Gamma$-$M$ that is weakly split by spin-orbit coupling due to only one metal atom per unit cell and the absence of interlayer splitting, which is consistent with the ARPES band dispersion shown in Fig. \ref{figure_2}i. The only other experimentally realized polytype with a single atom per unit cell is 1T-TaS$_2$. Since this phase has inversion symmetry, it would not show the spin-orbit coupling induced band splitting along the M-K direction that is clearly observed in Fig. \ref{figure_2}g. The observed Fermi-surface in Fig. \ref{figure_2}i must therefore originate from the 3R-TaS$_2$ phase.

It should also be noted that the valence band (as seen in Fig.\ref{figure_5}a and Fig.\ref{figure_5}b in the main text) shows at least two quantized levels. This suggests that the system is not a single 1H mono-layer.

\subsection{Spatial map}\label{supplementary_Spatial_map}
\begin{figure}[h]
    \centering
    \includegraphics[width = 0.8\textwidth]{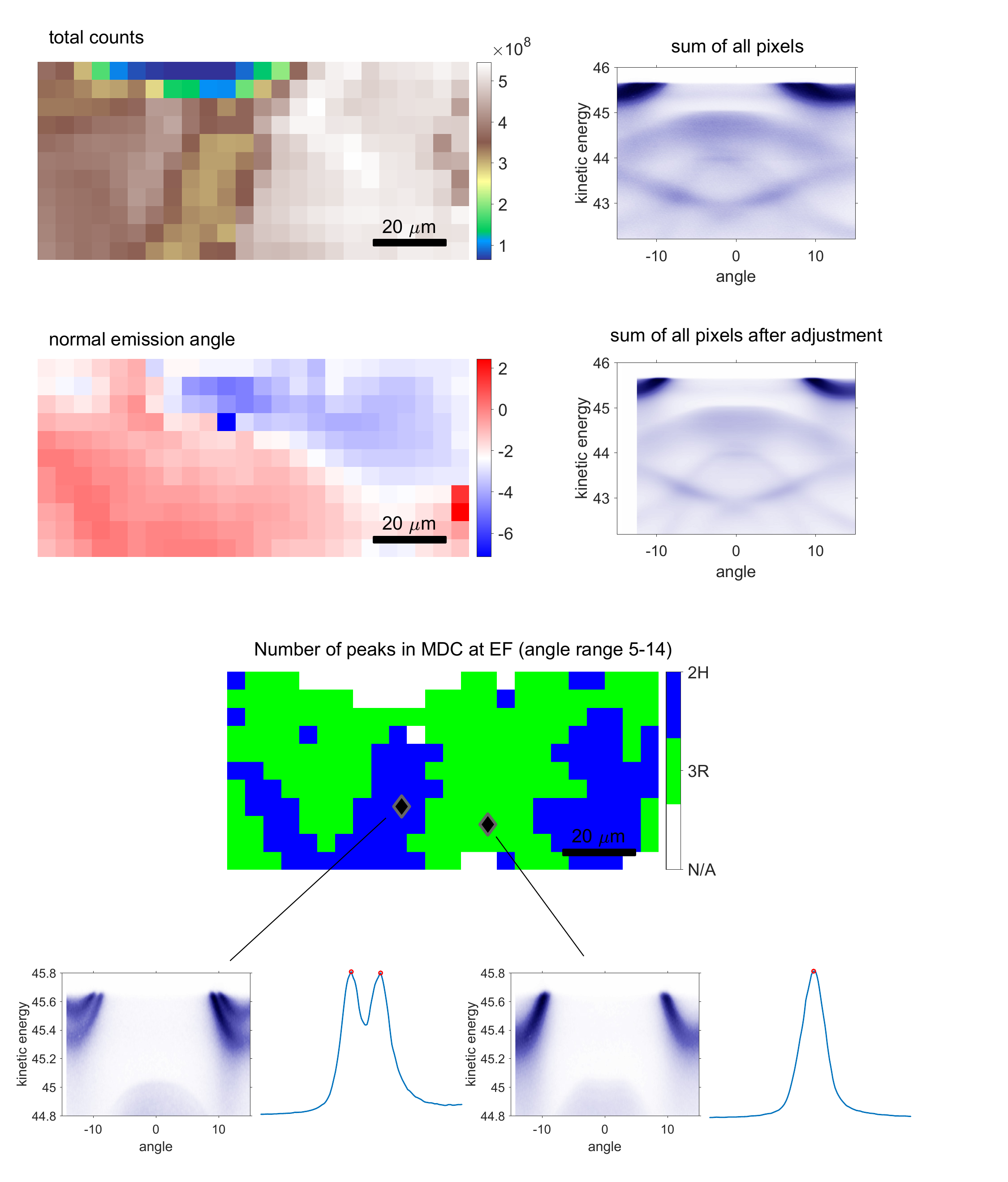}
    \caption{\textbf{Extracting domain of different polytypes from ARPES data.} Since TaS$_2$ is a flaky sample, the normal emission angle varies substantially in a raster scan of the sample. To account for this, we first apply
a routine to automatically determine the normal emission angle at each pixel of the spatial map, using an
autocorrelation function, and then adjust the angle scale of each pixel so normal emission is at 0. We then
extract MDCs in the angle range 5-14 degrees, and apply a peakfinding algorithm to the MDC. In case one
peak is found, we assign this pixel to the  3R phase, in case 2 peaks are found we assign the 2H phase, and in some regions where
mutiple peaks are found (due to flakes or noise or low counts) we assign N/A.}
    \label{figure_S_1}
\end{figure}

We use the distinction in the number of bands between 2H and 3R polytypes along the G-M direction to distinguish the two phases  when scanning the photon beam over the sample. This process is illustrated in Fig.~\ref{figure_S_1}.  

\subsection{Matrix elements effects}\label{supplementary_Matrix_elements}

\begin{figure}[h]
    \centering
    \includegraphics[width = 0.5\textwidth]{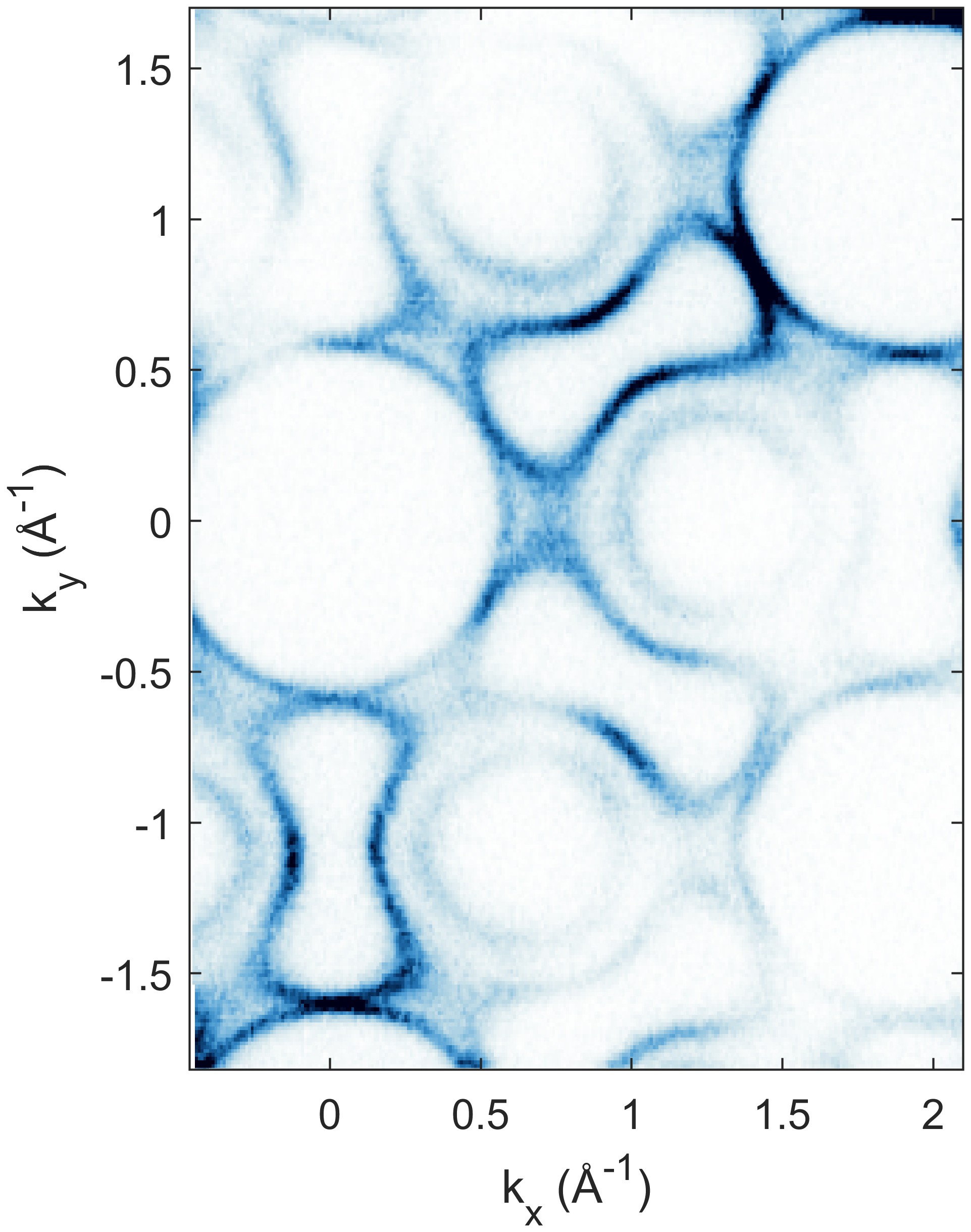}
    \caption{\textbf{Fermi surface map of the 3R phase measured with h$\nu =$ 126 eV photons.} We can see that even in the neighboring Brillouin zones, where the matrix elements are different, the hole and electron pockets are connected at one point.}
    \label{figure_S_matrix_elements}
\end{figure}

\end{document}